 \documentclass[smallabstract,smallcaptions]{dccpaper}

\usepackage{epsfig}
\usepackage{amsmath}
\usepackage{comment}
\usepackage{amssymb}
\usepackage{color}
\usepackage{url}

\newlength{\figurewidth}
\newlength{\smallfigurewidth}

\begin{document}

\title
{\large
\textbf{Compression of Higher Order Ambisonics with Multichannel RVQGAN}
}

\author{%
Toni Hirvonen and Mahmoud Namazi\\[0.5em]
{\small\begin{minipage}{\linewidth}\begin{center}
Samsung Research America \\
\url{[t.hirvonen, nima.namazi] @samsung.com}
\vspace*{0.2in}
\end{center}\end{minipage}}
}

\maketitle
\thispagestyle{empty}

\begin{abstract}
A multichannel extension to the RVQGAN neural coding method is proposed, 
and realized for data-driven compression of third-order Ambisonics audio. 
The input- and output layers of the generator and discriminator models are 
modified to accept multiple (16) channels without increasing the model bitrate. 
We also propose a loss function for accounting for spatial perception in 
immersive reproduction, and transfer learning from single-channel models. 
Listening test results with 7.1.4 immersive playback 
show that the proposed extension is suitable for coding 
scene-based, 16-channel Ambisonics content with good quality at 16 kbps when trained and tested on the EigenScape database. The model has potential applications for learning other types of content and multichannel formats.
\end{abstract}

\Section{Introduction}

Audio compression is an important aspect of storage and transport, in particular with respect to new immersive formats. The size and dimensionality of this type of  
content motivates the discovery of new, efficient coding methods. 
This is especially true in the case of Higher Order Ambisonics, a prominent surround sound methodology \cite{daniel2003further} 
which has in recent times been a popular solution especially for spatial sound capture.
At the same time, 
the rise of data-driven processing methods and increase in available computational power make it feasible to explore methods 
beyond conventional audio coding.

This paper presents a simple, yet general and efficient, multichannel extension to the popular 
RVQGAN audio coding methods suitable the compression of Ambisonics audio scenes. 
Despite the present paper being anchored in audio coding, i.e.\ faithfully reproducing 
an original reference, we also draw inspiration from music generation.
In addition to the modified model architecture, we also propose
customized training methods. As a novel testing aspect for neural codecs, 
multichannel loudspeaker listening is used to evaluate 
if the method is suitable 
for multichannel neural compression in terms of spatial quality and compression efficiency.

\Section{Background}

\SubSection{Neural Audio Coding and Music Generation}

In this section, we draw commonalities between neural coding and music generation, 
and survey how both fields relate to the present study. 

RVQGAN is a popular model for end-to-end neural audio coding \cite{Zeghidour2021SoundStreamAE, descript}.
Methods besides RVQ have been proposed, namely diffusion models \cite{Zhang2024HighFidelityDA, SanRoman2023FromDT} as well as 
transformer-based alternatives 
\cite{Gu2024ESCES}.
Neural codecs, despite having very good quality at particular low-bitrate range, have 
still suffer from many issues which impede real-life use. 
Complexity, especially on the decoder side, is an issue for real-life deployment on mobile devices.
End-to-end neural methods can have problems scaling towards complete transparency, hybrid systems 
address this by using generative neural techniques to enhance the quality of a traditional codec  \cite{dolbyaudiodecodinginverseproblem}.

The majority of neural audio coding research has been focused on single-channel aka mono audio compression. 
However, most modern audio content has more than one channel, making relevant the spatial and immersive aspects of audio. 
In recent decades, formats and reproduction systems utilizing more than two channels have emerged, starting from 5.1 surround and culminating with the immersive audio formats of recent years \cite{atmoscinema, AC-4, MPEG-H, iamf, AdSS}. 
Some attempts to apply neural coding methods to multichannel audio coding amount to processing a stereo input as dual-mono, thereby doubling the bitrate \cite{Defossez2022HighFN}. These works do not take advantage of the interchannel correlations present in the data. As an exception, Kleijn et al.\ \cite{kleijn_multichannelgan} generated multichannel (in the experiment only stereo) as a mono channel + spatial metadata, and  used generative methods for the spatial information. 
However, as opposed to being more generative in nature, the present study aims to replicate and compare to the original multichannel spatial impression.

For music generation, proposals have been oriented more toward stereo audio. 
\cite{copet2024simplecontrollablemusicgeneration, evans2024fasttimingconditionedlatentaudio, evans2024longformmusicgenerationlatent}
The reason is possibly that compression and faithful reproduction of the reference signal
is viewed as most suitable for content with more learnable structure than in 
unconstrained music, such as speech (which is typically mono). 
However, music generation uses looser measures of ``goodness'' that 
are not applicable to the coding paradigm. Also, as the generators typically 
have no audio input, not many insights can be obtained for multichannel encoding.

While acknowledging the transparency scaling issues and taking inspiration from the 
music generation studies, our goal is to remain within the end-to-end audio coding paradigm. 
We, however, limit this study to ambient background material that is typically 
more constrained in structure, and is a less critical part of the audio content 
in mixed presentations of multiple audio elements \cite{iamf}.
Testing other material and model complexity reduction is left as future work.

\SubSection{Ambisonics}

Ambisonics is a prominent audio representation widely utilized for spatial capture \cite{daniel2003further}.
This section briefly summarizes the technical aspects of the format. 
Higher Order Ambisonics (HOA) leverages spherical harmonics coefficients for sound field encoding. These coefficients, in conjunction with spherical harmonics as basis functions, are summed to approximate the original recorded sound field. Mathematically, for a given radius $r$, angle $\theta$, and wave number $k$, the pressure in a plane is represented by Equation \ref{eq:HOAeq}, where $B_{mm}^{\pm 1}$ denote the HOA coefficients.

\begin{equation}
\begin{aligned}
p(r,\theta) = & B_{00}^{+1}J_{0}(kr)  + \sum_{m=1}^{\infty}J_{m}(kr)B_{mm}^{+1}\sqrt{2}cos(m\theta) + \sum_{m=1}^{\infty}J_{m}(kr)B_{mm}^{-1}\sqrt{2}sin(m\theta)
\label{eq:HOAeq}
\end{aligned}
\end{equation}

To achieve faithful reproduction of the original sound field, 
theoretically, an infinite number of spherical harmonics would 
be needed. However, practical implementations necessitate 
truncation to a finite order $M$, where increased order yields 
enhanced fidelity. The resultant truncated multichannel HOA 
signal, known as B-format, demands $(M + 1)^2$ Ambisonics 
channels to represent the sound field accurately. For instance, 
first-order Ambisonics contains 4 channels, second-order 
Ambisonics contains 9 channels, and third-order Ambisonics 
contains 16 channels. Decoding these channels to speaker signals 
involves applying a decoding matrix contingent upon the HOA B-
format order and speaker positioning.

Despite its efficacy in capturing immersive audio environments, 
HOA poses challenges regarding data compression and 
transmission. The high channel count inherent in HOA entails 
substantial bitrate requirements for transmission. Thus, 
efficient compression schemes tailored to HOA data are 
imperative for practical deployment in various applications. Previous methods have relied on classical signal processing techniques involving linear combinations of lower-order channels or past channels \cite{Hellerud2009} or using techniques such as SVD to reduce the dimensionality of the Ambisonics signal \cite{Zamani2017, zamani2018spatial, Zamani2019, Mahe2019, xu2021higher, namazi2022spatial, namazi2024DCC, namazi2024ambisonics, Hold2024, NamaziThesis}. These compression methods are able to reduce bitrates from tens of megabits/second to the order of hundreds of kilobits/second. However, as proposed in this work, it may be possible to further reduce bitrates using a data-driven neural approach.

One mitigating aspect for real-life application 
may be that the material typically captured 
with HOA is more truncated in terms of probability distribution than 
unconstrained general audio. 
Modeling a general audio domain such as ``all music" is a hard problem, and 
many neural coding systems opt to test only a truncated distribution like speech
in order to get good quality. 
We propose that the typical use of HOA in transmission systems 
such as \cite{iamf} would also often involve a truncated distribution: 
spatial capture is often applied to background ambience and overall scene of e.g.\ alive event. Also, in modern transmission and content creation, the background can be 
complemented with separately coded discrete elements \cite{atmoscinema, iamf}.
Therefore, real-life application of the neural methods for scene-based audio 
can benefit from the low rates offered by data-driven compression, 
and can be viable even without scaling to high-quality transparency.

\Section{Methods}

\SubSection{Model Architecture}

The basic idea of the present multichannel architecture extension can be summarized as simply 
increasing the channel count of the first and last convolutional layer of RVQGAN 
to match the number of audio channels. In the case of 3rd-order Ambisonics material, 
this channel count is 16. 
With such a change, it is possible to keep the dimensionality, and therefore the compression efficiency of the model bottleneck.
Besides straightforward architectural changes, 
this implies non-trivial changes to the typical loss functions (see next section).

The standard convolution layer used in DNNs \cite{dumoulin2018guideconvolutionarithmeticdeep} utilizes multiple channels and kernels, which can be also thought of as FIR filters. 
Time-domain PCM audio signals are typically treated as 1-D time series signals, i.e.\ 
the kernels are one-dimensional. In this case, the layer output value 
with input size $(C_{in}, L)$ and output $(C_{out}, L_{out})$ is

\begin{equation}
    \text{out}(C_{out}) = \text{bias}(C_{out}) + \sum_{k=0}^{C_{in}-1} \text{kernel}(C_{out}, k) \mbox{*} \text{input}(k)
    \label{eq:conv}
\end{equation}

\noindent where \mbox{*} is the valid cross-correlation operator, 
C denotes a number of channels and L the length of the signal sequence.
By using the standard arithmetic, the operation each channel of the original audio is 
processed with their own dedicated kernel tailored to the specifics of that signal,
and the results are summed to the next layer signal. The kernel filters are optimized to 
both find the inter-channel structure, and to best represent the linear combination of channels 
in the output sequences. In previous neural codecs with monaural input, the first layer 
rather only performs an upmixing operation. 

In addition to a brute-force solution of running a monaural model separately in each channel, 
an alternative method could be to capture the inter-channel structure via 
two-dimensional kernels (as in 2-D convolutional layers). 
However, this would increase the computational complexity, 
and introduce many separate notions of channels in the model, which are arguably more difficult to interpret. 

Justifications for the present solution of using convolutional arithmetic to handle
inter-channel aspects of the audio can be found in perceptual spatial hearing studies \cite{spatialhearing}. Much of auditory localization can be explained by interaural 
time- and level differences, and the overall spatial envolopment is tied to interaural correlation \cite{spatialhearing}.
In non-neural codecs, these aspects can be represented by preserving the inter-channel 
covariance structure \cite{AC-4, MPEG-H}.
Also, spatial attributes can be quantized heavily, and are typically assigned less bitrate compared to waveform coding, and can be more generative in nature.
We will discuss the signal covariance structure more in the next section 
in relation to loss functions.

\SubSection{Loss Functions}
\label{sec:loss}

The original Descript RVQGAN model was trained with the combination of adversarial-, feature matching-, VQ codebook-, and reconstruction losses \cite{descript}. 
Since our model output is multichannel, we adjust 
the relevant losses to operate on each channel independently and finally take the expected value over channels as the final loss. 
For the adversarial discriminator, this requires to change the 
multi-scale (MSD) and multi-period and waveform discriminators (MPD), as well as 
the multi-resolution spectrogram
discriminators (MRSD) to produce multichannel output.

We found that despite not looking at the channel interactions
specifically, the previous paradigm already produces a 
good spatial impression for the present HOA content.
For fine-tuning of spatial quality, we consider a loss for the mismatch of the 
interchannel covariance structure, 
which has been found to be strongly related to perceived interaural coherence, and therefore an important descriptor of perceptual spatial impression or the rendered output \cite{spatialhearing, blauert_ic, interauralcoh}. 
Attributes this loss aims to preserve are e.g.\ envelopment and diffuseness of the material.
For content where these are important to emphasize, the covariance loss is a simple way to do that.
It is also more general than the specialized loss functions proposed for stereo audio \cite{evans2024fasttimingconditionedlatentaudio, steinmetz2020automaticmultitrackmixingdifferentiable}.

The covariance loss is calculated as the expected L1 loss between the normalized 
channel-wise covariance matrices
of the original and the model $f_\theta$ reconstruction time-domain signals:

\begin{equation}
    L_{cov} = \frac{1}{2}\sum_{i=0}^n\sum_{j=0}^n||\frac{C_{ij}}{\sqrt{C_{ii}C_{jj}}}-\frac{\hat{C}_{ij}}{\sqrt{\hat{C}_{ii}\hat{C}_{jj}}}||, 
    \label{eq:covloss}
\end{equation}

\noindent where 
$C=\text{cov}(\mathbf{x})$ and $\hat{C}=\text{cov}(f_\theta(\mathbf{x}))$ are the covariance matrices of 
the multichannel time-domain input and model output signal of $n$ channels, respectively.
Each element of the normalized matrix is then the Pearson correlation coefficient 
between channels $i$ and $j$.


It has been found that such preservation of the convariance structure between channels 
is a useful target in spatial coding systems, and also account for possible linear 
rendering of the output afterward \cite{AC-4, MPEG-H, engdegrd2010mpeg}.
We evaluate the covariance broadband, but it is also possible 
to have the measure be frequency-dependent. 
We also experimented using non-normalized covariance instead of correlation coefficient, 
but found the latter more appealing for numerical stability, and the fact that 
the other losses seemed to also account for channel energies and level differences. 

\SubSection{Transfer Learning from Single-Channel Models} \label{TransferLearning}

We propose applying transfer learning to the
multichannel model from previously trained single-channel model weights. 
In input and output layers, where the number of channels is increased, 
we replicate the original convolutional weights of the original channel to 
the multiple channels of the new model identically. 
Thus, all input and output channels 
of the multichannel model are processed with identical convolutional kernels at
the start of the training, which then evolve as needed. 

\begin{figure}[t]
\begin{center}
\epsfig{width=3.5in,file=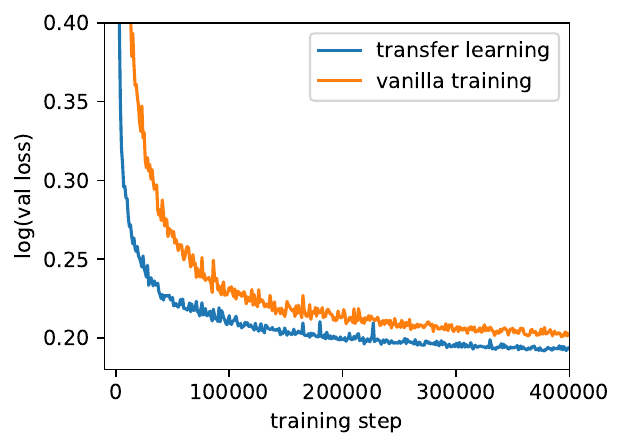}
\end{center}
\caption{\label{fig:transfer}%
Example comparison between proposed transfer learning utilizing pre-trained single-channel model weights, and vanilla random initialization. 
Plot shows the development of the reconstruction multi-scale mel validation loss  \cite{descript}.}
\end{figure}

In practice, we construct the new model and copy the appropriate parameters 
from the state dictionary of the published 16 kbps version of 
Descript monaural model \cite{descript}.
In addition to helping with RVQGAN training time consumption, 
transfer learning also helps if the fine-tuning stage is performed on 
less powerful hardware and smaller batch size.
Fig.\ \ref{fig:transfer} illustrates the benefit of the proposed 
transfer learning in both faster conversion, and final error with 
reasonable amount of steps. 

In our experiments, we do not apply transfer learning the adversarial discriminator model, only randomly initialize it as in the original work \cite{descript}.

\SubSection{Experiment}

Immersive audio material available for data-driven modeling is not as abundant as traditional formats. In this paper, we have utilized the EigenScape database \cite{eigenscape}.
It consists of eight acoustic scenes recorded spatially in 4th-order Ambisonics that 
are ambient in nature, but do sometimes include e.g.\ people speaking. We omit the additional channels and process the signals to obtain 16-channel, 3rd-order inputs at 16-bit / 44100 Hz resolution. 
For cross-validation, we utilize 7/8 of the samples on each scene for 
model training, and the keep the rest separate for validation and subjective testing.
A multiple-round alternative of keeping each of the 8 individual scenes 
for validation, and testing them in turn was unfortunately 
not feasible due to subjective testing being used. However, as we use transfer learning from a model trained with generic audio, the resulting model should retain some level of baseline generality.

The model implementation was built on the Descript codebase \cite{descript} 
with the aforementioned multichannel extensions to the architecture and loss calculation. 
The only other changes were to omit the output Tanh-nonlinearity, and trivial modifications to make the dependencies work with 16-channel audio. 
The training procedure remains largely the same as
in the original work, while we utilize the transfer learning scheme proposed 
in the previous section. 
We use weighting of 1.0 for the covariance loss. The other loss weightings 
were kept as is (15.0 for the multi-scale mel loss, 2.0 for the feature
matching loss, 1.0 for the adversarial loss and 1.0, 0.25 for the codebook and commitment losses, respectively). 
To illustrate a simple fine tuning procedure results, 
we trained for a modest number of steps (400000) and batch size (24) 
of 5-second samples. 

In order to evaluate and compare the proposed Ambisonics neural codec, a MUSHRA \cite{MUSHRA} 
listening test was carried out. 
For quality reference, we utilize
Opus channel mapping family 3 coding intended for HOA content \cite{skoglund2018ambisonics}, at 160 kbps rate, 
which is on the lower quality region of the Opus HOA, to demonstrate 
how conventional coding methods with 10x bitrate compare.
In total, four listening conditions were compared:

\begin{itemize}
    \item Low anchor (LA) - 3.5kHz low-pass filtered version of the reference
    \item Opus' Channel Family Mapping 3 (OPUS) compression at 160 kbps
    \item Proposed neural ambisonics codec (PROP) at 16 kbps 
    \item Hidden reference (H-REF)
\end{itemize}

Rendering the HOA signals to the 7.1.4 loudspeaker layout was done with the IAMF system \cite{iamf} which utilizes a version of the EBU ADM Renderer \cite{itu2127}.
Listening tests took place in a 7m (L) x 5.33m (W) x 3.05 (H) listening room equipped with a 7.1.4 playback system. The loudspeakers' and listener's positions were based on \cite{itu2051}. Loudspeakers were level-matched at the listener's position and meet the ITU-R specification \cite{itu1116}.

Eight listeners completed the MUSHRA test. Participants were told to compare the four listening conditions to the known reference, and to rate the overall sound quality on the MUSHRA 100-point scale, while focusing on the audio quality and spatial impression correctness versus the hidden reference.

\Section{Results and Discussion}

\begin{figure}[h]
\begin{center}
\epsfig{width=5.5in,file=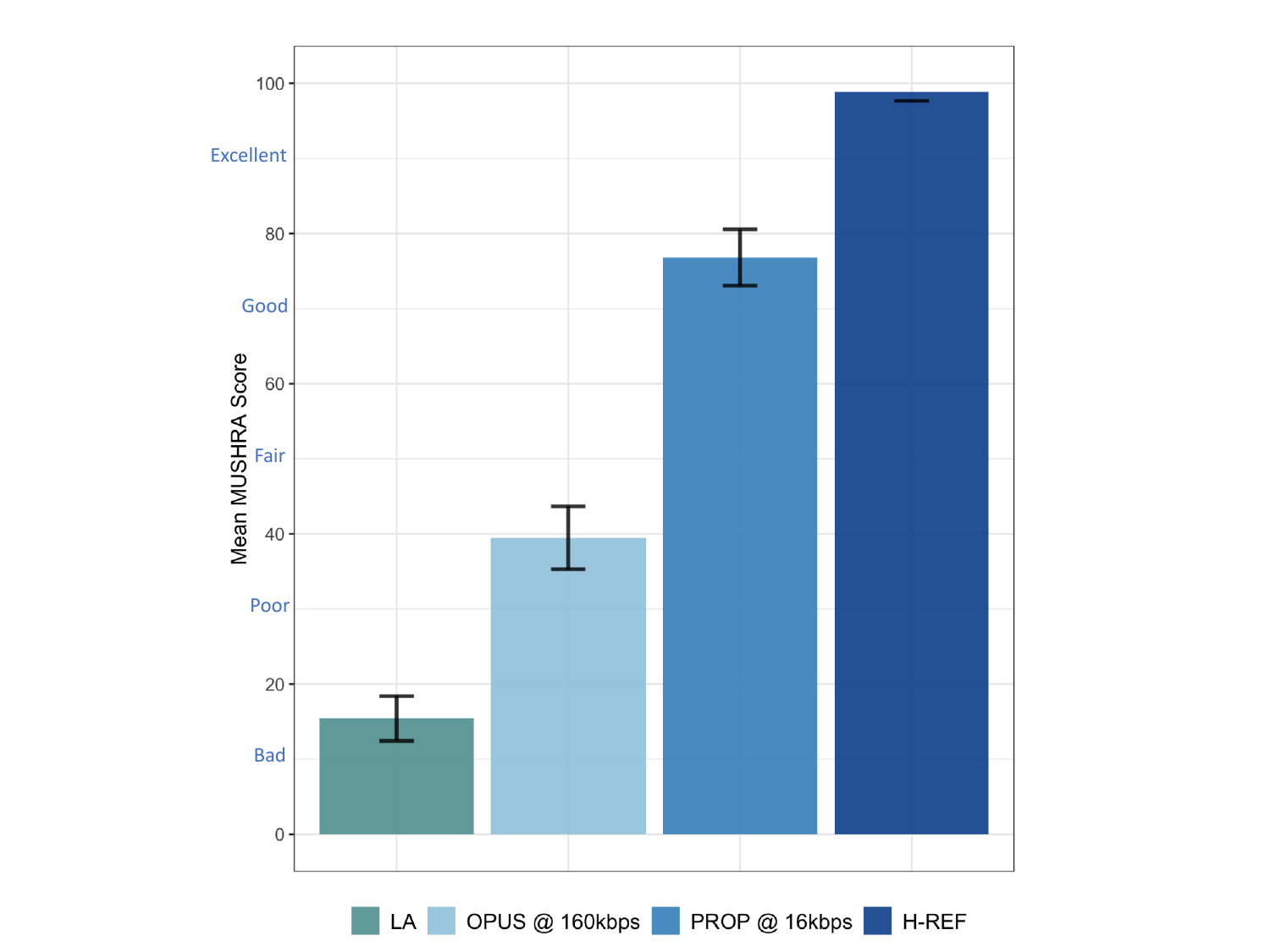}
\end{center}
\caption{\label{fig:results}%
MUSHRA score mean and 95\% confidence interval over 8 tracks and 8 listeners, 
with 7.1.4 immersive loudspeaker listening.
}
\end{figure}

As can be seen from Fig.\ \ref{fig:results}, the proposed neural codec operating 
on 16 kbps is able to achieve "good" quality on the MUSHRA scale, and outperform 
a traditional method with 10x less compression rate. While the traditional channel family mapping Opus method 
starts to break down at 160 kbps, and its quality would improve at higher rates, 
we opted to illustrate the quality potential of the neural methods rather than 
find the precise rate where the two methods would match. 

The audio material used here is from a truncated distribution of ambient scenes, and also limited in amount in order to illustrate a typical use case, as discussed in
the Ambisonics-section earlier. 
However, the use of transfer learning from a generic model arguably retains 
some level of baseline generality.
Also, the listening test conditions did enable critical evaluation of the 
sound quality similarly as in typical immersive audio testing. 
Although the present material was ambient background in nature, we did not test mixed presentations with additional foreground audio elements \cite{iamf}, which would have made the HOA coding quality less critical. 

As the main caveat, it should be noted that informal listening with few musical samples outside the EigenScape database indicated that applying the proposed model trained only on scene-based material may not be as suitable for all content. 
Other content types or more generic models can also require increased bitrate to achieve good quality. 
More general training and application of the model with e.g.\ musical and cinematic content is left for future.

Although not tested here, 
the present modeling principle is also applicable to multichannel audio formats
other than 3rd-order Ambisonics (e.g.\ 5.1, 7.1, 7.1.4 etc.), as long as there is sufficient 
training data. Despite the promising results, 
the main caveats of the present method, and most neural codecs remain
1) the lack of scaling to different bitrates without retraining or model modification, 
and 2) the model computational complexity. For these, future work will hopefully
provide solutions.

\Section{Acknowledgements}

The authors would like to thank Samsung’s US Audio Lab
staff and especially Ema Souza Blanes for helping with the listening tests.

\Section{References}
\bibliographystyle{IEEEbib}
\bibliography{refs}

\begin{thebibliography}{10}

\bibitem{daniel2003further}
Jerome Daniel, Sebastien Moreau, and Rozenn Nicol,
\newblock ``Further investigations of high-order ambisonics and wavefield synthesis for holophonic sound imaging,''
\newblock in {\em Audio Engineering Society Convention 114}, March 2003.

\bibitem{Zeghidour2021SoundStreamAE}
Neil Zeghidour, Alejandro Luebs, Ahmed Omran, Jan Skoglund, and Marco Tagliasacchi,
\newblock ``Sound{S}tream: An end-to-end neural audio codec,''
\newblock {\em IEEE/ACM Transactions on Audio, Speech, and Language Processing}, vol. 30, pp. 495--507, 2021.

\bibitem{descript}
Rithesh Kumar, Prem Seetharaman, Alejandro Luebs, Ishaan Kumar, and Kundan Kumar,
\newblock ``High-fidelity audio compression with improved {RVQGAN},'' 2023,
\newblock arXiv 2306.06546.

\bibitem{Zhang2024HighFidelityDA}
Zhengpu Zhang, Jianyuan Feng, Yongjian Mao, Yehang Zhu, Junjie Shi, Xuzhou Ye, Shilei Liu, Derong Liu, and Chuanzeng Huang,
\newblock ``High-fidelity diffusion-based audio codec,''
\newblock {\em 2024 18th International Workshop on Acoustic Signal Enhancement (IWAENC)}, pp. 344--348, 2024.

\bibitem{SanRoman2023FromDT}
Robin~San Roman, Yossi Adi, Antoine Deleforge, Romain Serizel, Gabriel Synnaeve, and Alexandre D'efossez,
\newblock ``From discrete tokens to high-fidelity audio using multi-band diffusion,'' 2023,
\newblock arXiv 2308.02560.

\bibitem{Gu2024ESCES}
Yuzhe Gu and Enmao Diao,
\newblock ``{ESC}: Efficient speech coding with cross-scale residual vector quantized transformers,'' 2024,
\newblock arXiv 2404.19441.

\bibitem{dolbyaudiodecodinginverseproblem}
Pedro J.~Villasana T., Lars Villemoes, Janusz Klejsa, and Per Hedelin,
\newblock ``Audio decoding by inverse problem solving,'' 2024,
\newblock arXiv 2409.07858.

\bibitem{atmoscinema}
Charles Robinson, Nicholas Tsingos, and Shripal Mehta,
\newblock ``Scalable format and tools to extend the possibilities of cinema audio,''
\newblock {\em SMPTE Motion Imaging Journal}, vol. 121, no. 8, November 2012.

\bibitem{AC-4}
``{Digital Audio Compression (AC-4) Standard},''
\newblock Standard, European Telecommunications Standards Institute, Geneva, CH, 2018.

\bibitem{MPEG-H}
``{Information technology - High efficiency coding and media delivery in heterogeneous environments - Part 3: 3D audio},''
\newblock Standard, International Organization for Standardization, Geneva, CH, 2022.

\bibitem{iamf}
``Immersive {A}udio {M}odel and {F}ormats,'' \url{https://aomedia.org/iamf/}, 2024.

\bibitem{AdSS}
``{Advanced sound system for programme production},''
\newblock Standard, International Telecommunication Union, Geneva, CH, 2022.

\bibitem{Defossez2022HighFN}
Alexandre D'efossez, Jade Copet, Gabriel Synnaeve, and Yossi Adi,
\newblock ``High fidelity neural audio compression,'' 2022,
\newblock arXiv 2210.13438.

\bibitem{kleijn_multichannelgan}
W.~Bastiaan Kleijn, Michael Chinen, Felicia S.~C. Lim, and Jan Skoglund,
\newblock ``Multi-channel audio signal generation,''
\newblock in {\em ICASSP 2023 - 2023 IEEE International Conference on Acoustics, Speech and Signal Processing (ICASSP)}, 2023, pp. 1--5.

\bibitem{copet2024simplecontrollablemusicgeneration}
Jade Copet, Felix Kreuk, Itai Gat, Tal Remez, David Kant, Gabriel Synnaeve, Yossi Adi, and Alexandre Défossez,
\newblock ``Simple and controllable music generation,'' 2024,
\newblock arXiv 2306.05284.

\bibitem{evans2024fasttimingconditionedlatentaudio}
Zach Evans, CJ~Carr, Josiah Taylor, Scott~H. Hawley, and Jordi Pons,
\newblock ``Fast timing-conditioned latent audio diffusion,'' 2024,
\newblock arXiv 2402.04825.

\bibitem{evans2024longformmusicgenerationlatent}
Zach Evans, Julian~D. Parker, CJ~Carr, Zack Zukowski, Josiah Taylor, and Jordi Pons,
\newblock ``Long-form music generation with latent diffusion,'' 2024,
\newblock arXiv 2404.10301.

\bibitem{Hellerud2009}
Erik Hellerud, Audun Solvang, and U.~Peter Svensson,
\newblock ``Spatial redundancy in higher order ambisonics and its use for lowdelay lossless compression,''
\newblock in {\em 2009 IEEE International Conference on Acoustics, Speech and Signal Processing}, 2009, pp. 269--272.

\bibitem{Zamani2017}
Sina Zamani, Tejaswi Nanjundaswamy, and Kenneth Rose,
\newblock ``Frequency domain singular value decomposition for efficient spatial audio coding,''
\newblock in {\em 2017 IEEE Workshop on Applications of Signal Processing to Audio and Acoustics (WASPAA)}, 2017, pp. 126--130.

\bibitem{zamani2018spatial}
Sina Zamani and Kenneth Rose,
\newblock ``Spatial audio coding with backward-adaptive singular value decomposition,''
\newblock in {\em Audio Engineering Society Convention 145}. Audio Engineering Society, 2018.

\bibitem{Zamani2019}
Sina Zamani and Kenneth Rose,
\newblock ``Spatial audio coding without recourse to background signal compression,''
\newblock in {\em ICASSP 2019 - 2019 IEEE International Conference on Acoustics, Speech and Signal Processing (ICASSP)}, 2019, pp. 720--724.

\bibitem{Mahe2019}
Pierre Mah{\'e}, Stephane Ragot, and Sylvain Marchand,
\newblock ``{First-order ambisonic coding with quaternion-based interpolation of PCA rotation matrices},''
\newblock in {\em {EAA Spatial Audio Signal Processing Symposium}}, Paris, France, Sept. 2019, pp. 7--12.

\bibitem{xu2021higher}
Jiahao Xu, Yadong Niu, Xihong Wu, and Tianshu Qu,
\newblock ``Higher order ambisonics compression method based on independent component analysis,''
\newblock in {\em Audio Engineering Society Convention 150}. Audio Engineering Society, 2021.

\bibitem{namazi2022spatial}
Mahmoud Namazi, Ahmed Elshafiy, and Kenneth Rose,
\newblock ``Spatial audio compression with adaptive singular value decomposition using reconstructed frames,''
\newblock in {\em Audio Engineering Society Conference: 2022 AES International Conference on Audio for Virtual and Augmented Reality}. Audio Engineering Society, 2022.

\bibitem{namazi2024DCC}
Mahmoud Namazi, Ahmed Elshafiy, and Kenneth Rose,
\newblock ``On ultra low-delay compression of higher order ambisonics signals,''
\newblock in {\em 2024 Data Compression Conference (DCC)}, 2024, pp. 512--521.

\bibitem{namazi2024ambisonics}
Mahmoud Namazi and Kenneth Rose,
\newblock ``Ultra-low delay lossless compression of higher order ambisonics,''
\newblock in {\em ICASSP 2024 - 2024 IEEE International Conference on Acoustics, Speech and Signal Processing (ICASSP)}, 2024, pp. 791--795.

\bibitem{Hold2024}
Christoph Hold, Leo McCormack, Archontis Politis, and Ville Pulkki,
\newblock ``Perceptually-motivated spatial audio codec for higher-order ambisonics compression,''
\newblock in {\em ICASSP 2024 - 2024 IEEE International Conference on Acoustics, Speech and Signal Processing (ICASSP)}, 2024, pp. 1121--1125.

\bibitem{NamaziThesis}
Mahmoud Namazi,
\newblock {\em Advancements in Higher Order Ambisonics Compression and Loss Concealment Techniques},
\newblock Ph.D. thesis, UC Santa Barbara, 2024.

\bibitem{dumoulin2018guideconvolutionarithmeticdeep}
Vincent Dumoulin and Francesco Visin,
\newblock ``A guide to convolution arithmetic for deep learning,'' 2018,
\newblock arXiv 1603.07285.

\bibitem{spatialhearing}
Jens Blauert,
\newblock ``Spatial hearing: The psychophysics of human sound localization (revised edition),''
\newblock 1997.

\bibitem{blauert_ic}
Jens Blauert and Werner Lindemann,
\newblock ``{Spatial mapping of intracranial auditory events for various degrees of interaural coherence},''
\newblock {\em The Journal of the Acoustical Society of America}, vol. 79, no. 3, pp. 806--813, March 1986.

\bibitem{interauralcoh}
Robert Luke, Hamish Innes-Brown, Jaime~A Undurraga, and David Mcalpine,
\newblock ``Human cortical processing of interaural coherence,''
\newblock {\em iScience}, vol. 5, no. 25, March 2022.

\bibitem{steinmetz2020automaticmultitrackmixingdifferentiable}
Christian~J. Steinmetz, Jordi Pons, Santiago Pascual, and Joan Serrà,
\newblock ``Automatic multitrack mixing with a differentiable mixing console of neural audio effects,'' 2020,
\newblock arXiv 2010.10291.

\bibitem{engdegrd2010mpeg}
Jonas Engdegård et~al.,
\newblock ``{MPEG} spatial audio object coding - the {ISO}/{MPEG} standard for efficient coding of interactive audio scenes,''
\newblock in {\em AES 129th Convention}. Audio Engineering Society, 2010.

\bibitem{eigenscape}
Marc~Ciufo Green and Damian Murphy,
\newblock ``Eigenscape,'' \url{https://doi.org/10.5281/zenodo.1012809}, 2017.

\bibitem{MUSHRA}
ITU-R,
\newblock ``Method for the subjective assessment of intermediate quality level of audio systems,''
\newblock Recommendation BS.1534-2, International Telecommunication Union, Geneva, 2014.

\bibitem{skoglund2018ambisonics}
Jan Skoglund and Michael Graczyk,
\newblock ``{A}mbisonics in an {O}gg {O}pus container,'' \url{https://www.rfc-editor.org/rfc/rfc8486.html}, 2018.

\bibitem{itu2127}
ITU-R,
\newblock ``Audio definition model renderer for advanced sound systems,''
\newblock Recommendation BS.2127-1, International Telecommunication Union, Geneva, 2023.

\bibitem{itu2051}
ITU-R,
\newblock ``Advanced sound system for programme production,''
\newblock Recommendation BS.2051-3, International Telecommunication Union, Geneva, 2022.

\bibitem{itu1116}
ITU-R,
\newblock ``Methods for the subjective assessment of small impairments in audio systems,''
\newblock Recommendation BS.1116-3, International Telecommunication Union, Geneva, 2015.

\end{thebibliography}

\end{document}